\documentclass{article}
\def\rd{{\rm d}}

\begin{document}

\title{\bf Actions for Vacuum Einstein's Equation with a Killing Symmetry}
\author{Han He, Yongge Ma\thanks{Email address: ygma@163.com}, and Xuejun Yang\\
\small Department of Physics, Beijing Normal University, Beijing
100875, CHINA}
\date{\today}
\maketitle

\begin{abstract}
In a space-time $M$ with a Killing vector field $\xi^a$ which is
either everywhere timelike or everywhere spacelike, the collection
of all trajectories of $\xi^a$ gives a 3-dimensional space $S$.
Besides the symmetry-reduced action from that of Einstein-Hilbert,
an alternative action of the fields on $S$ is also proposed which
gives the same fields equations as those reduced from the vacuum
Einstein equation on $M$.

PACS number(s): 04.20.Fy, 04.20.Cv

\end{abstract}

\section{Introduction}

Dimensional reduction is a crucial step in any high dimensional
theory of physics such as, Kaluza-Klein theory \cite{Duff_PR_130}
and string theory \cite{Green_Cambridge_1987}, in order to make
contact with the 4-dimensional sensational world. It is also a
useful approach to study 4-dimensional space-time with symmetries
in general relativity.

In the appendix of \cite{Geroch_JMP_12}, Geroch introduced a
Killing reduction formalism of 4-dimensional space-time. Let
$(M,g_{ab})$ be a space-time with a Killing vector field $\xi^a$,
and $\xi^a$ is either everywhere timelike or everywhere spacelike.
The collection of all trajectories of $\xi^a$ gives a
3-dimensional space $S$. Geroch's discussion shows that there is a
one-to-one correspondence between tensor fields and tensor
operations on $S$ and certain tensor fields and tensor operations
on $M$. Thus, the differential geometry of $S$ will be mirrored in
$M$. Geroch gives the curvature tensor and geometric properties of
$S$ and finally the field equations on $S$, which is equivalent to
the vacuum Einstein equation on $M$. If the Killing vector field
$\xi^a$ is hypersurface orthogonal, there is a well-known
conclusion that $3+1$ gravity is equivalent to $2+1$ gravity
coupled to a massless scalar field.

Geroch's work concerned only the reduction of the equation of
motion. In this paper, we will study the Killing reduction from
the viewpoint of variation principle. It turns out that, if one
starts with the symmetry-reduced Hilbert action, the 4-metric
components rather than the variables of Geroch have to be regarded
as the arguments in order to get the correct reduced fields
equations. An alternative action in the arguments of Geroch's
3-dimensional fields is also obtained, which gives the same field
equations as Geroch's on $S$. If we confined the configuration to
static space-time, both actions are reduced to that of
3-dimensional Euclidean gravity coupled to a massless Klein-Gordon
field, which is consistent with the reduced action in
\cite{Ma_PRD_65}.

\section{Killing reduction of space-time}

Let $(M,g_{ab})$ be a space-time with Killing vector field
$\xi^a$, which is either everywhere timelike or everywhere
spacelike. Let $S$ denote the collection of all trajectories of
$\xi^a$. An element of $S$ is an inextendible curve in $M$ which
is everywhere tangent to $\xi^a$.

A mapping $\psi$ from $M$ to $S$ can be defined as follows: For
each point $p$ of $M$, $\psi(p)$ is the trajectory of $\xi^a$
passing through $p$. Assume $S$ is given the structure of a
differentiable 3-manifold such that $\psi$ is a smooth mapping.

If the Killing field $\xi^a$ were hypersurface orthogonal, it
would be possible to represent $S$ as one of the hypersurfaces in
$M$ which is everywhere orthogonal to $\xi^a$. Each trajectory of
$\xi^a$ would intersect this hypersurface in exactly one point.

In the non-hypersurface-orthogonal case. It is most natural to
regard $S$ as a quotient space of $M$ rather than a subspace.
Geroch shows that there is a one-to-one correspondence between
tensor fields $T'^{b\cdots d}_{a\cdots c}$ on $S$ and tensor
fields $T^{b\cdots d}_{a\cdots c}$ on $M$ which satisfy
\[
  \xi^a T^{b\cdots d}_{a\cdots c}=0,\ \cdots \ ,\xi_d T^{b\cdots d}_{a\cdots
  c}=0 ,
\]
\begin{equation}\label{relations_S_M}
  {\cal{L}}_\xi T^{b\cdots d}_{a\cdots c}=0 .
\end{equation}
The entire tensor algebra on $S$ is completely and uniquely
mirrored by tensors on $M$ subject to (\ref{relations_S_M}). So
the primes will be dropped hereafter: Speaking of tensor fields
being on $S$ is merely as a shorthand way of saying that the field
on $M$ satisfies (\ref{relations_S_M}).

The metric and inverse metric on $S$ are defined as
\begin{equation}
  h_{ab}=g_{ab}-\lambda^{-1}\xi_a\xi_b,
\end{equation}
\begin{equation}
  h^{ab}=g^{ab}-\lambda^{-1}\xi^a\xi^b,
\end{equation}
\begin{equation}
  h_{a}^{b}=\delta_{a}^{b}-\lambda^{-1}\xi_a\xi^b,
\end{equation}
where $\lambda$ is the norm of Killing field $\xi^a$, i.e.,
\begin{equation}
  \lambda=\xi^m\xi_m.
\end{equation}
Note that the indices of any tensor on $S$ can be raised or
lowered with either $h$ or $g$ with the same result.

The covariant derivative operator on $S$ is defined as
\begin{equation}
  D_e T^{b\cdots d}_{a\cdots c}=h^m_a\cdots h^n_c\ h^b_r\cdots h^d_s\
  h^p_e \nabla_p T^{r\cdots s}_{m\cdots n},
\end{equation}
which satisfies all the conditions of derivative operator and
\begin{equation}
  D_c h_{ab}=0.
\end{equation}

The twist of $\xi^a$ is defined by
\begin{equation}
  \omega_a=\epsilon_{abcd}\xi^b\nabla^c\xi^d.
\end{equation}
The Ricci tensor ${\mathcal{R}}_{ab}$ of $h_{ab}$ on $S$ is
related to that of $g_{ab}$ on $M$ by
\begin{eqnarray}\label{Ricci_relation}
  {\mathcal{R}}_{ab}&=&\frac{1}{2}\lambda^{-2}(\omega_a\omega_b-h_{ab}
  \omega^c\omega_c)+\frac{1}{2}\lambda^{-1}D_a D_b \lambda \nonumber\\
  &&-\frac{1}{4}\lambda^{-2}(D_a \lambda)(D_b \lambda)+h^c_ah^d_bR_{cd}.
\end{eqnarray}
Contracting Eq. (\ref{Ricci_relation}) we get the relation between
the scalar curvatures ${\mathcal{R}}$ of $h_{ab}$ and $R$ of
$g_{ab}$ as
\begin{eqnarray}\label{scalar_curvature}
  {\mathcal{R}}&=&-\frac{1}{2}\lambda^{-2}\omega_a\omega_bh^{ab}
  +\lambda^{-1}h^{ab}D_a D_b \lambda \nonumber\\
  &&-\frac{1}{2}\lambda^{-2}h^{ab}(D_a \lambda)(D_b \lambda)+R.
\end{eqnarray}
As shown in \cite{Geroch_JMP_12}, in the source-free case (the
4-dimensional Ricci tensor $R_{ab}=0$), $\omega_a$ is a gradient:
\begin{equation}
  \omega_a=D_a\omega.
\end{equation}
The Ricci tensor ${\mathcal{R}}_{ab}$ on $S$ then takes the form
\begin{eqnarray}\label{Ricci_relation_no_source}
  {\mathcal{R}}_{ab}&=&\frac{1}{2}\lambda^{-2}[(D_a\omega)(D_b\omega)-h_{ab}
  (D^m\omega)(D_m\omega)] \nonumber\\
  &&+\frac{1}{2}\lambda^{-1}D_a D_b \lambda
  -\frac{1}{4}\lambda^{-2}(D_a \lambda)(D_b \lambda).
\end{eqnarray}
and the scalar fields on $S$ satisfy
\begin{equation}\label{D_2_lambda_Geroch}
  D^2\lambda=\frac{1}{2}\lambda^{-1}(D^m\lambda)(D_m\lambda)-\lambda^{-1}
  (D^m\omega)(D_m\omega),
\end{equation}
\begin{equation}\label{D_2_omega_Geroch}
  D^2\omega=\frac{3}{2}\lambda^{-1}(D^m\lambda)(D_m\omega).
\end{equation}
Eqs. (\ref{Ricci_relation_no_source}), (\ref{D_2_lambda_Geroch})
and (\ref{D_2_omega_Geroch}) are equivalent to the vaccum
Einstein's equation of $g_{ab}$.

\section{Alternative actions on $S$}

\subsection{The symmetry-reduced Hilbert action}

The Einstein-Hilbert action on $M$ is defined by
\begin{equation}\label{E-H_action_for_M}
  S_{EH}[g^{ab}]=\int_M{\mathcal{L}}_G=\int_M\sqrt{-g}R,
\end{equation}
where g is the determinant of 4-metric components $g_{\mu\nu}$ in
some basis on $M$.

Suppose the Killing vector field is everywhere timelike. For
practical calculations, it is convenient to take a coordinate
system adapted to the congruence:
\begin{equation}
  t^a=\left(\frac{\partial}{\partial t}\right)^a=\xi^a.
\end{equation}
Then the line element of $g_{ab}$ on $M$ can be written as
\cite{Kramer_Cambridge_1980}
\begin{equation}
  \rd s^2=h_{\mu\nu}\rd x^\mu\rd x^\nu+\lambda(\rd t+B_\mu\rd
  x^\mu)^2.
\end{equation}
The components of metric are
\begin{equation}
  g_{\mu\nu}=h_{\mu\nu}+\lambda^{-1}\xi_\mu\xi_\nu,\ \ g_{\mu 4}=\xi_\mu=\lambda B_\mu,\ \
  g_{44}=\xi_4=\lambda .
\end{equation}
We also have
\begin{equation}\label{metric component}
  2B_{[\mu,\nu]}=|\lambda|^{-3/2}\epsilon_{\mu\nu\rho}\omega^{\rho},\ \ \epsilon_{\mu\nu\rho}=|\lambda|^{-1/2}
  \epsilon_{\mu\nu\rho\sigma}\xi^\sigma .
\end{equation}
Let $h$ denote the determinant of $h_{\mu\nu}$ on $S$.
Straightforward calculations give
\begin{eqnarray}\label{relation_between_g_h}
  g = \lambda h .
\end{eqnarray}
This formula is still valid in the case where $\xi^a$ is
everywhere spacelike.

Since the principle of symmetric criticality is valid in the one
Killing vector model of general relativity \cite{FelsCQG}, one
expects that the 3-dimensional fields Eqs.
(\ref{Ricci_relation_no_source}), (\ref{D_2_lambda_Geroch}) and
(\ref{D_2_omega_Geroch}) could be obtained by the variations of
the action from symmetric reduction of (\ref{E-H_action_for_M}),
which can be read from Eq. (\ref{scalar_curvature}) as
\begin{eqnarray}\label{E-H_action_for_S}
  S_{EH}=\int_S|\lambda|^{1/2}\sqrt{|h|}({\mathcal{R}}+\frac{1}{2}\lambda^{-2}\omega_a\omega_bh^{ab}),
\end{eqnarray}
where a 3-dimensional total divergence term has been neglected.
This reduced action cannot lead to the correct reduced fields
equations by variations with respect to the Geroch's variables
$(h^{ab}, \lambda, \omega_a)$. For instance, the variation of
(\ref{E-H_action_for_S}) with respect to $\omega_a$ leads to
$\omega_a=0$, which is not the case in general. This is not a very
surprising result since the relation of $\omega_a$ and the
4-metric components is rather nontrivial, as one can see from Eq.
(\ref{metric component}). Although the variation of Hilbert action
with respect to the 4-metric gives certainly the Einstein
equation, there is no guarantee to get the same equation if one
varies it with respect to other variables arbitrarily concocted
such as $\omega_a$. Eq. (\ref{metric component}) suggests a
one-form $A_a$ at least locally on $S$ such that
\begin{equation}\label{transform_relation_of A}
(dA)_{ab}=|\lambda|^{-3/2}\epsilon_{bac}\omega^c.
\end{equation}
Substituting $A_a$ for $\omega_a$ in Eq. (\ref{E-H_action_for_S}),
we get the familiar form of the symmetry-reduced action
\begin{eqnarray}\label{E-H_action_of A for_S}
  S_{EH}[h^{ab},A_c,\lambda]=\int_S|\lambda|^{1/2}\sqrt{|h|}[{\mathcal{R}}-\frac{1}{4}\lambda
  h^{ab}h^{cd}(dA)_{ac}(dA)_{bd}].
\end{eqnarray}
The variations of this action with respect to
$(h^{ab},A_c,\lambda)$ will give the correct reduced fields
equations on $S$, that can be checked directly by substituting
$\omega_a$ back for $A_a$ after variations.

\subsection{Hypersurface orthogonal case}

We now consider a special case where the Killing field $\xi^a$ is
hypersurface orthogonal. It is then possible to represent $S$ as
one of the hypersurfaces in $M$ which is everywhere orthogonal to
$\xi^a$. The symmetry-reduced Einstein-Hilbert action on $S$ reads
\begin{eqnarray}\label{E-H_action_for_S_before_transform_orthogonal}
  S_{EH}[h^{ab}]=\int_S\mathcal{L}_G
  =\int_S|\lambda|^{1/2}\sqrt{|h|}{\mathcal{R}}.
\end{eqnarray}

To obtain regular form of action on $S$, we now conformally
transform $h_{ab}$ as $\tilde{h}_{ab}$:
\begin{equation}\label{transform_relation_of h}
  \tilde{h}_{ab}=\Omega^{-2}h_{ab}.
\end{equation}
Let $\tilde{D}_a$ be the covariant derivative operator determined
by metric $\tilde{h}_{ab}$, i.e.,
\begin{equation}
  \tilde{D}_a \tilde{h}_{bc}=0.
\end{equation}
$\tilde{D}_a$ is related to $D_a$ by
\begin{equation}\label{transform_relation_of D}
  D_a v_b = \tilde{D}_a v_b - C^c_{ab}v_c,
\end{equation}
where
\begin{eqnarray}\label{transform_relation_of C}
  C^c_{ab}&=&\frac{1}{2}h^{cd}(\tilde{D}_a h_{bd}+\tilde{D}_b h_{ad}-\tilde{D}_d h_{ab}) \nonumber \\
  &=&\tilde{h}^c_b\tilde{D}_a\ln\Omega+\tilde{h}^c_a\tilde{D}_b\ln\Omega
      -\tilde{h}_{ab}\tilde{h}^{cd}\tilde{D}_d\ln\Omega.
\end{eqnarray}
After the conformal transform, the form of action
(\ref{E-H_action_for_S_before_transform_orthogonal}) becomes
\begin{eqnarray}\label{E-H_action_for_S_after_transform_orthogonal}
  S_{EH}[\tilde{h}^{ab}]
  &=&\int_S|\lambda|^{1/2}\sqrt{\Omega^6|\tilde{h}|}\ \Omega^{-2}\Big[\tilde{\mathcal{R}}-4\tilde{h}^{ab}\tilde{D}_a\tilde{D}_b\ln\Omega \nonumber \\
  &&-2\tilde{h}^{ab}(\tilde{D}_a\ln\Omega)(\tilde{D}_b\ln\Omega)\Big].
\end{eqnarray}
Let
\begin{equation}\label{relation between Omega_and lambda}
  \Omega=|\lambda|^{-1/2},
\end{equation}
then (\ref{E-H_action_for_S_after_transform_orthogonal}) becomes
\begin{eqnarray}\label{E-H_action_for_S_after_transform_orthogonal_1}
  S_{EH}[\tilde{h}^{ab}]
  =\int_S\sqrt{|\tilde{h}|} \Big[\tilde{\mathcal{R}}-4\tilde{h}^{ab}\tilde{D}_a\tilde{D}_b\ln\Omega
  -2\tilde{h}^{ab}(\tilde{D}_a\ln\Omega)(\tilde{D}_b\ln\Omega)\Big].
\end{eqnarray}
Note that the second term in
(\ref{E-H_action_for_S_after_transform_orthogonal_1}) is a total
divergence term that can be ignored for variations. Let
\begin{equation}
\Lambda\equiv\sqrt{2}\ln\Omega,
\end{equation}
straightforward calculations of variations show that the action
(\ref{E-H_action_for_S_after_transform_orthogonal_1}) gives the
same equations of motion as those of 3-gravity $\tilde{h}^{ab}$
coupled to a massless Klein-Gordon field $\Lambda$, which is
defined by the coupled action:
\begin{equation}\label{E-H_action_for_S_after_transform_orthogonal_2}
  S_E+S_{KG}=\int_{S}\sqrt{|\tilde{h}|}[\tilde{R}-\tilde{h}^{ab}
  (\partial_a\Lambda)(\partial_b\Lambda)].
\end{equation}
Therefore, a 4-dimensional space-time with a hypersurface
orthogonal Killing vector field which is either everywhere
timelike or everywhere spacelike, is ``conformally'' equivalent to
3-dimensional gravity coupled to a massless scalar field. Static
space-time is of course a typical case \cite{Ma_PRD_65}.

\subsection{An alternative action}

In Geroch's reduction, the 4-dimensional Einstein vacuum
$(M,g_{ab})$ reduces in general to a 3-dimensional gravity coupled
to two scalar fields on $S$ satisfying Eqs.
(\ref{Ricci_relation_no_source}), (\ref{D_2_lambda_Geroch}) and
(\ref{D_2_omega_Geroch}). One may ask if there is an action in the
arguments of Geroch's variables $\omega$ and $\lambda$ on $S$,
whose variations give the above Geroch's equations. Note that in
hypersurface orthogonal case, besides $\tilde{h}_{ab}$ only one
scalar field $\lambda$ exists on $S$. Up to a boundary term the
action (\ref{E-H_action_for_S_after_transform_orthogonal_1}) can
be written as
\begin{eqnarray}\label{action orthogonal}
  S_{EH}[\tilde{h}^{ab}]
  =\int_S\sqrt{|\tilde{h}|} \Big[\tilde{\mathcal{R}}
  -\frac{1}{2}\lambda^{-2}\tilde{h}^{ab}(\tilde{D}_a\lambda)(\tilde{D}_b\lambda)\Big].
\end{eqnarray}
While in general case, another scalar field $\omega$ exists on
$S$. By carefully observing Geroch's equations and the special
action (\ref{action orthogonal}), we propose an action for the
reduced fields equations on $S$ as
\begin{eqnarray}\label{E-H_action_for_S_after_transform}
  S[\tilde{h}^{ab}]
  =\int_S\sqrt{|\tilde{h}|} \Big[\tilde{\mathcal{R}}
  -\frac{1}{2}\lambda^{-2}\tilde{h}^{ab}(\tilde{D}_a\lambda)(\tilde{D}_b\lambda)
  -\frac{1}{2}\lambda^{-2}\tilde{h}^{ab}(\tilde{D}_a\omega)(\tilde{D}_b\omega)\Big].
\end{eqnarray}
We now show that the variations of
(\ref{E-H_action_for_S_after_transform}) give exactly the same
field equations as those of Geroch's work.

The variation of (\ref{E-H_action_for_S_after_transform}) with
respect to $\tilde{h}^{ab}$ gives
\begin{eqnarray}\label{action_bian_fen_second}
  \delta S &=&\int_S \sqrt{|\tilde{h}|}\
  \bigg[
    \tilde{\mathcal{R}}_{ab}-\frac{1}{2}\tilde{\mathcal{R}}\tilde{h}_{ab}
   +\frac{1}{4}\lambda^{-2}\tilde{h}_{ab}\tilde{h}^{cd}(\tilde{D}_c\lambda)(\tilde{D}_d\lambda)-\frac{1}{2}\lambda^{-2}(\tilde{D}_a\lambda)(\tilde{D}_b\lambda) \nonumber \\
   &&\quad\quad\quad\quad +\frac{1}{4}\lambda^{-2}\tilde{h}_{ab}\tilde{h}^{cd}(\tilde{D}_c\omega)(\tilde{D}_d\omega)-\frac{1}{2}\lambda^{-2}(\tilde{D}_a\omega)(\tilde{D}_b\omega)
  \bigg]\delta\tilde{h}^{ab}.
\end{eqnarray}
The vanishing of (\ref{action_bian_fen_second}) gives the Einstein
field equation of gravity with sources on $S$ as:
\begin{eqnarray}\label{Einstein_field_equation_on_S}
  \tilde{\mathcal{R}}_{ab}-\frac{1}{2}\tilde{\mathcal{R}}\tilde{h}_{ab}
   &=&\frac{1}{2}\lambda^{-2}(\tilde{D}_a\lambda)(\tilde{D}_b\lambda)-\frac{1}{4}\lambda^{-2}\tilde{h}_{ab}\tilde{h}^{cd}(\tilde{D}_c\lambda)(\tilde{D}_d\lambda) \nonumber \\
   &&
   +\frac{1}{2}\lambda^{-2}(\tilde{D}_a\omega)(\tilde{D}_b\omega)-\frac{1}{4}\lambda^{-2}\tilde{h}_{ab}\tilde{h}^{cd}(\tilde{D}_c\omega)(\tilde{D}_d\omega),
\end{eqnarray}
or
\begin{equation}\label{Ricci_curvature_after_transform}
   \tilde{\mathcal{R}}_{ab}
   =\frac{1}{2}\lambda^{-2}(\tilde{D}_a\lambda)(\tilde{D}_b\lambda)
   +\frac{1}{2}\lambda^{-2}(\tilde{D}_a\omega)(\tilde{D}_b\omega).
\end{equation}

Likewise the variation principle gives the equations of motion of
the matter fields $\lambda$ and $\omega$ determined by action
(\ref{E-H_action_for_S_after_transform}) respectively as:
\begin{equation}\label{coupled_field_equation_after_transform}
  \tilde{h}^{ab}\tilde{D}_a\tilde{D}_b\lambda=\lambda^{-1}\tilde{h}^{ab}(\tilde{D}_a\lambda)(\tilde{D}_b\lambda)
   -\lambda^{-1}\tilde{h}^{ab}(\tilde{D}_a\omega)(\tilde{D}_b\omega),
\end{equation}
and
\begin{equation}\label{coupled_field_equation_after_transform_2}
  \tilde{h}^{ab}\tilde{D}_a\tilde{D}_b\omega
      =2\lambda^{-1}\tilde{h}^{ab}(\tilde{D}_a\lambda)(\tilde{D}_b\omega).
\end{equation}

To compare with Geroch's work \cite{Geroch_JMP_12}, we make the
conformal transformation (\ref{transform_relation_of h})
inversely. From (\ref{transform_relation_of D}),
(\ref{transform_relation_of C}) and (\ref{relation between
Omega_and lambda}), we have
\begin{eqnarray}\label{transform_relation_of_D_square_lambda}
  \tilde{h}^{ab}\tilde{D}_a\tilde{D}_b\lambda &=& \Omega^2\left[h^{ab}D_a D_b\lambda
      -h^{ab}(D_a \lambda) (D_b\ln\Omega)\right] \nonumber \\
  &=& \Omega^2\left[D^2\lambda+\frac{1}{2}\lambda^{-1}h^{ab}(D_a
      \lambda)(D_b\lambda)\right],
\end{eqnarray}
and
\begin{eqnarray}\label{transform_relation_of_D_square_omega}
  \tilde{h}^{ab}\tilde{D}_a\tilde{D}_b\omega &=& \Omega^2\left[h^{ab}D_a
      D_b\omega
      -h^{ab}(D_a \omega) (D_b\ln\Omega)\right] \nonumber \\
  &=& \Omega^2\left[D^2\omega+\frac{1}{2}\lambda^{-1}h^{ab}(D_a
      \omega)(D_b\lambda)\right].
\end{eqnarray}

Substituting (\ref{transform_relation_of_D_square_lambda}) into
(\ref{coupled_field_equation_after_transform}), one gets equation
(\ref{D_2_lambda_Geroch}). Substituting
(\ref{transform_relation_of_D_square_omega}) into
(\ref{coupled_field_equation_after_transform_2}), one gets
equation (\ref{D_2_omega_Geroch}).

The relation between $\mathcal{R}_{ab}$ and
$\tilde{\mathcal{R}}_{ab}$ reads (see the appendix D of
\cite{Wald_Chicago_1984})
\begin{eqnarray}\label{relation_between_R_ab}
  \mathcal{R}_{ab} &=& \tilde{\mathcal{R}}_{ab}-D_a
     D_b\ln\Omega-h_{ab}h^{cd}D_c D_d\ln\Omega \nonumber \\
  &&
  -(D_a\ln\Omega)(D_b\ln\Omega)+h_{ab}h^{cd}(D_c\ln\Omega)(D_d\ln\Omega).
\end{eqnarray}
Substituting (\ref{relation_between_R_ab}) into
(\ref{Ricci_curvature_after_transform}), we obtain
\begin{eqnarray}
  \mathcal{R}_{ab} &=& -\frac{1}{4}\lambda^{-2}(D_a\lambda)(D_b\lambda)
     +\frac{1}{2}\lambda^{-2}(D_a\omega)(D_b\omega) \nonumber \\
  && +\frac{1}{2}\lambda^{-1}D_a D_b\lambda
     -\frac{1}{4}\lambda^{-2}h_{ab}(D^m\lambda)(D_m\lambda)
     +\frac{1}{2}\lambda^{-1}h_{ab}D^2\lambda.
\end{eqnarray}
Using (\ref{D_2_lambda_Geroch}), one gets equation
(\ref{Ricci_relation_no_source}).

In conclusion, the variations of the symmetry-reduced Hilbert
action (\ref{E-H_action_for_S}) with respect to Geroch's variables
on $S$ cannot lead to the correct reduced field equations. While,
the variations of action (\ref{E-H_action_for_S_after_transform})
defined on $S$ give the field equations for 3-dimensional gravity
coupled to two scalar fields, which are ``conformally'' equivalent
to Geroch's equations reduced from the vacuum Einstein equation on
$M$. In this sense, we may regard the action
(\ref{E-H_action_for_S_after_transform}) as an alternative action
for vacuum Einstein's equation with a Killing symmetry. Geroch's
equations can thus be obtained from the variation principle.

The above discussion could be generalized to higher dimensional
case \cite{Yang_preperation}, and hence might lead some new sight
on the mechanism of dimensional reduction.

\section*{Acknowledgments}

This work is supported in part by NSFC No. 10205002. H. He and X.
Yang would like to acknowledge support from NSFC(10073002). Y. Ma
would also like to acknowledge support from Young Teachers
Foundation of BNU and YSRF for ROCS, SEM.

\end{document}